\documentclass[aps,prd,showpacs,amsmath,amssymb,superscriptaddress,nofootinbib,onecolumn,10pt]{revtex4-1}
\usepackage[]{latexsym}
\usepackage[english]{babel}
\usepackage{graphicx}
\usepackage{hyperref}
\usepackage{color}

\begin{document}

\title{Emerging the dark sector from thermodynamics of cosmological systems with constant pressure}

\author{Alejandro Aviles}
\email{aviles@ciencias.unam.mx}
\affiliation{Departamento de Matem\'aticas, Cinvestav del Instituto Polit\'ecnico Nacional (IPN), 07360, M\'exico DF, Mexico.}

\author{Norman Cruz}
\email{norman.cruz@usach.cl}
\affiliation{Departamento de F\'isica, Universidad de Santiago de Chile, Av.~Ecuador 3493, Santiago, Chile.}

\author{Jaime Klapp}
\email{jaime.klapp@inin.gob.mx}
\affiliation{Departamento de Matem\'aticas, Cinvestav del Instituto Polit\'ecnico Nacional (IPN), 07360, M\'exico DF, Mexico.}
\affiliation{Departamento de F\'{\i}sica,  Instituto Nacional  de Investigaciones Nucleares, Apartado Postal, 18-1027,  11801, M\'exico DF, M\'exico.}

\author{Orlando Luongo}
\email{luongo@na.infn.it}
\affiliation{Dipartimento di  Fisica, Universit\`a di Napoli ``Federico II", Via Cinthia, I-80126, Napoli, Italy.}
\affiliation{Istituto Nazionale di Fisica Nucleare (INFN), Sezione di Napoli, Via Cinthia, I-80126 Napoli, Italy.}
\affiliation{Instituto de Ciencias Nucleares, Universidad Nacional Aut\'onoma de M\'exico (UNAM), M\'exico, DF 04510, Mexico.}

\date{\today}

\begin{abstract}
We investigate the thermodynamics of general fluids that have the constriction that their pressure is constant. 
For example, this happens in the case of pure {\it dust} matter, for which the pressure vanishes and also in the 
case of standard dark matter phenomenology. Assuming a finite non-zero pressure, the corresponding dynamics is richer 
than one naively would expect. In particular, it can be considered as a \emph{unified description} of dark energy and 
dark matter. We first consider the more general thermodynamic properties of this class of fluids finding the important 
result that for them adiabatic and isothermal processes should coincide. We therefore study their behaviors in curved 
space-times where \emph{local thermal equilibrium} can be appealed. Thus, we show that this \emph{dark fluid} 
degenerates with the dark sector of the $\Lambda$CDM model only in the case of adiabatic evolution. We demonstrate that, 
adding dissipative processes, a phantom behavior can occur and finally we further highlight that an arbitrary decomposition 
of the dark sector, into \emph{ad hoc} dark matter and dark energy terms, may give rise to phantom dark energy, whereas the
whole dark sector remains non-phantom.
\end{abstract}

\maketitle


\section{Introduction}

The most accepted cosmological picture suggests that the universe is
mostly filled by \emph{dark species}, which comprise about the $96\%$ of
the total energy content \cite{CervantesCota:2011pn,Copeland:2006wr}.
The dark species are usually decomposed into two fluids: dark matter and
dark energy. Dark matter seems to clump at all scales,
being responsible for the formation of gravitational potential wells in
which baryons fall, forming structures we observe. Conversely, dark energy
provides a negative pressure which counteracts the gravitational
attraction, accelerating the universe today. The two species, dubbed
\emph{the universe dark sector}, are apparently not related to each other
and so one supposes both dark energy and dark matter to be two completely
separated constituents. However, there is no unequivocal
evidence that the stress-energy tensor of the total dark sector should be
split into those two components, leading to a degeneracy problem between
cosmological species \cite{Kunz:2007rk}. Afterwards, there is no reason to
assume the dark sector is formed by a single component
\cite{Hu:1998tj,Aviles:2011ak,LUONGO:2014haa,Xu:2011bp,Bielefeld:2014gca}
or by many counterparts, which in principle may also interact
\cite{Aviles:2011ak}.
The consequence of this \emph{dark degeneracy} leads to a \emph{plethora} of
theoretical paradigms, each of them indistinguishable from the standard
$\Lambda$CDM model. This degeneracy can be broken, for example, by assuming
that dark matter is the clustering component of the dark sector.
This serves as a dark matter definition and leaves unsolved the degeneracy problem itself.
Indeed, if the dark sector does not interact at all with the particle standard model, it is impossible to \emph{disentangle} every
single dark constituent. Even in presence of interactions, it is not clearly understood
how the degeneracy definitively breaks down \cite{Aviles:2011ak}. An attractive
description of the dark sector is obtained by employing one single
fluid with vanishing speed of sound
\cite{Hu:1998tj,Aviles:2011ak,LUONGO:2014haa,Xu:2011bp}.
This feature ensures fluid perturbations to grow up at all scales
\cite{Aviles:2011ak} and also provides a negative pressure that is
responsible for the universe speeding up \cite{Aviles:2011ak,LUONGO:2014haa}. This fluid
is sometimes also named the {\it dark fluid}. One of the advantages of
introducing the dark fluid is that dark matter and dark energy are
considered as \emph{emergent features} of a single equation of state (EoS). This
is also the underlying philosophy of the generalized Chaplygin gas
\cite{Kamenshchik:2001cp,Bento:2002ps,Bilic:2001cg}, which lies on the
fundamental advantage that its description can be formulated in terms of
scalar fields. Further, it is worth mentioning the Chaplygin gas may also
arise from a dark matter viscous fluid \cite{Fabris:2005ts}, or more
generally from purely dissipative effects
\cite{Zimdahl:2000zm,Avelino:2008ph,Li:2009mf,Avelino:2010pb,Piattella:2011bs}. Coming back to perfect barotropic fluids with vanishing speed of sound,
it is easy to notice that their descriptions are equivalent to consider a perfect fluid with
constant pressure.

In this paper, we investigate the thermodynamical properties of such a class of fluids.
In so doing, we take into account dissipative processes which are typically
associated to real fluids. Their introduction is well motivated, due to
the fact that generic fluids are usually non-perfect and there exists no
robust reasons to characterize the whole dark sector with perfect fluids
only. Hence, we depart from thermodynamical equilibrium evolution, by
considering either the Eckart \cite{Eckart:19403} or causal Muller-Israel-Stewart (MIS)
theories
\cite{Muller1967,Israel:1976tn,Israel:1979wp,Pavon1982,Hiscock:1983zz},
showing moreover that our general results do not depend on which approach is
employed. Our hypothesis points out to interpret the dark fluid as an
approximate description of the dark sector in thermodynamic equilibrium.
In our single-fluid approach, both cold dark matter and non-evolving dark
energy essentially arise as special cases of zero pressure and zero
enthalpy respectively\footnote{We will clarify later that standard dark
energy fluids turn out to be unstable when dissipative effects are involved.}.
Thus, we highlight how a standard thermodynamic treatment may open new
insights on the study of the dark sector, showing a general landscape to
determine its EoS. We are also supported by recent developments on
dark energy relativistic thermodynamics which have been investigated both in
equilibrium and in the presence of dissipative processes, see {\it e.g.}
\cite{Lima:2004wf,GonzalezDiaz:2004eu,Pereira:2008af,Myung:2009km,Silva:2011pc,Aviles:2012et,Luongo:2012dv,Silva:2013ixa}.
Another interesting point is our study on the standard $\Lambda$CDM
phenomenology, at equilibrium and out-of-equilibrium, also  investigating
the role of degeneracy. To do so, we compare evolving dark energy terms at
both equilibrium and non-equilibrium stages. For example, if one assumes
as evolving dark energy the $w$CDM model, in which the dark energy term is
proportional to $(1+z)^{3(1+w)}$, with $w$ a constant EoS parameter, almost all measurements of $w$ are consistent with a non-evolving dark energy
\cite{Komatsu:2010fb,Ade:2013zuv,Chuang:2013wga},
nevertheless there is a trend in its expected value to be $w<-1$,
leading to phantom dark energy
\cite{Caldwell:1999ew,Caldwell:2003vq}. Phantom dark energy violates the
dominant energy condition \cite{Carroll:2003st}, and consequently there
exist observers measuring a negative energy density. To account this
strange behavior, several possibilities occur: one of them lies on accepting the
existence of negative energy densities in nature, whereas a second is to
abandon the pure $w$CDM model with $w<-1$. A relevant fact is that there
exists also the possibility that phantom behavior arises as a dissipative
process in the dark energy sector, as shown in
\cite{Barrow:2004xh,Cataldo:2005qh,Cruz:2006ck}. It is important to notice
that the phantom fluid violates the
the dominant energy condition, while a unified fluid does not necessarily violate it. That is why the dark sector decomposition may not exist at all,
providing an appealing alternative to the standard cosmological approach. In fact, considering, for example, a
unified fluid with EoS $P_u=w_{u}\rho_u$, and decomposing it as $\rho_u =
\rho_{ph} + \rho_{dm}$, where {\it ph} stands for phantom and
$dm$ for dark matter, with EoS parameters
$w_{ph}=-\epsilon < -1$ and $w_{dm} = \epsilon \rho_{ph} /\rho_{dm} + w_u
(\rho_{ph} + \rho_{dm})/\rho_{dm}$, it is evident
that the background evolution of both descriptions are the same, without
necessarily violating the dominant energy condition. We accurately show this
property in the case of our unified dark fluid. For those reasons, one of
the main purposes of this article is to investigate thermodynamical
systems, leading to the same phenomenology of the $\Lambda$CDM model,
but with profound physical departures from it, as out-of-equilibrium processes are
involved.

The paper is organized as follows. In Sec.~\ref{DarkFluidRev} we review
several  properties of the dark fluid. In Sec.~\ref{Section:general} we
analyze
the general properties of a system in thermodynamical equilibrium with
constant pressure. In Sec.~\ref{Section:BackCosm} we study the evolution
of these systems in a FRW background.
In Sec.~\ref{Sect:EntProd} we depart from thermodynamical equilibrium by considering the
entropy production due to dissipative processes.
We study the corresponding universe evolution and we show how out-of-equilibrium
processes may lead the dark fluid to become phantom.
We also summarize the consequence of first order perturbed equations in the field of cosmological perturbation theory. 
Finally in Sec.~\ref{Sect:Conclusions} we present our conclusions and perspectives.


\section{The cosmological standard model and the dark fluid}
\label{DarkFluidRev}

The \emph{dark fluid} definition may be formulated in terms of either perfect
fluids or ideal gases. In the perfect fluid picture, one employs a barotropic
EoS providing a vanishing adiabatic speed of sound at all
stages
of the universe evolution \cite{Aviles:2011ak}. Since the need of
adiabatic perturbations,
providing a vanishing sound speed, is not stringent, one can define the
barotropic condition even without postulating the EoS at the
beginning.\footnote{In \cite{Hu:1998tj} the barotropic condition is not considered and
the EoS of the dark fluid in a FRW universe is imposed from the beginning.}
The case of perfect gas with vanishing speed of sound \cite{LUONGO:2014haa}
needs no considerations on gravitational instabilities. Indeed, one argues that for
barotropic fluids, gravitational instability is driven
by the competition between gravitational attraction and pressure. Under
the barotropic condition, the zero sound speed allows fluid perturbations to grow at all
scales, as cold dark matter actually does. We can write the EoS of a
barotropic fluid without losing generality as
\begin{equation}
 P_d=w_d(\rho_d)\rho_d\,,
\end{equation}
where the subscript $d$ stands for $dark$. From the restriction of vanishing sound speed,
$c^2_s = dP_d/ d\rho_d = 0$, one obtains for the dark fluid
\begin{equation}
 w_d = \frac{P_0}{\rho_d}\,,
\end{equation}
which implies the pressure to be constant, i.e. $P_d=P_0$. If the dark fluid
is entirely
constituted by cold dark matter, the net pressure does not contribute
since it is expected to vanish. Astronomical observations, however, do not
exclude \emph{a priori} a  non-zero pressure,
permitting the existence of a small pressure, even of the order of
universe's critical density today, $\rho_c \equiv 3H^2_0/8\pi G$.

For example, a recent analysis of rotation curves in
low surface brightness galaxies has shown that $|w_{dm}| < 10^{-6}$ at galaxy centers
\cite{Barranco:2013wy}.
This indication enables one to imagine the cosmic speed up to be due to a
non-vanishing
dark fluid's pressure. Hence, the possibility that dark energy and dark
matter emerge as
a \emph{single effect} due to the dark fluid is plausible.
To understand how this is possible, let us consider the continuity
equation in a Friedman-Robertson-Walker (FRW) space-time
\begin{equation} \label{DarkFluidRev:AdGibbs}
 \dot{\rho}_d + 3 \frac{\dot{a}}{a} (\rho_d + P_0) = 0\,.
\end{equation}
This equation can be integrated to give
\begin{equation} \label{DarkFluidRev:rhodEv}
 \rho_d(a) = \frac{\rho_{d\,0}}{1+ \mathcal{K}} \left( 1 +
\frac{\mathcal{K}}{a^3} \right)\,,
\end{equation}
where $\mathcal{K} = - (\rho_{d0} + P_0)/P_0$ is an integration constant,
fitted considering $\rho_{d0}$ as the value of the dark fluid energy
density
evaluated at $a(t_0) \equiv a_0 = 1 $. We can notice that Eq.~(\ref{DarkFluidRev:rhodEv}) is what one expects for
a unified dark sector fluid. It is the sum of a component that decays as $\propto a^3$ and a
component which remains constant.

To guarantee a positive energy density at all epochs, $\mathcal{K}$ must be
positive and therefore the pressure turns out to be negative, lying within
the interval
$-\rho_{d0}\le P_0 \le 0$. This agrees with current time bounds on cosmic
acceleration
and allows the dark fluid to accelerate the universe.
Eq.~(\ref{DarkFluidRev:rhodEv}) definitively forecasts that the dark fluid
reproduces the same phenomenology of the $\Lambda$CDM model at the level of \emph{background cosmology}.

\noindent Thereby, the EoS parameter of the dark fluid can be written as
\begin{equation}
 w_d(a) = -\frac{1}{1+\mathcal{K}a^{-3}},
\end{equation}
which may be compared to the $\Lambda$CDM case, consisting of a
cosmological constant and cold dark matter contribution
\begin{equation} \label{wLCDM}
 w_{\Lambda + \text{dm}}(a) =
-\frac{1}{1+\frac{\Omega_{\text{dm}}}{\Omega_\Lambda}a^{-3}}\,,
\end{equation}
where $\Omega_\Lambda = 8 \pi G \rho_{\Lambda,0}/H_0^2$.
By comparing the two above equations, the correspondence among the two models is attained through the identifications
\begin{equation} \label{DarkFluidRev:DFLCDMId}
 \mathcal{K} = \frac{\Omega_{\text{dm}}}{\Omega_\Lambda},
 \qquad \text{and} \qquad \Omega_{d} = \Omega_{\text{dm}} + \Omega_\Lambda\,.
\end{equation}
This leads to a degeneracy
problem which persists even at linear orders of cosmological perturbation
theory\footnote{See \cite{Aviles:2011ak} for an explicit demonstration.}.
Despite of this degeneracy, we notice an important physical difference
between the $\Lambda$CDM and dark fluid paradigms: in the
$\Lambda$CDM model to describe the observed late-time acceleration,  it is
necessary to include the cosmological constant
which is commonly reinterpreted as a quantum field vacuum energy
contribution. This identification
leads to the well-known \emph{cosmological constant problem}, probably one
of the most serious inconsistencies between theoretical physics and observations
\cite{Weinberg:1988cp}. The dark fluid does not suffer from this shortcoming. In fact, in Eq.~(\ref{DarkFluidRev:rhodEv})
the term $\mathcal{K}$ does not contain any cosmological information associated to vacuum energy,
although the problem to physically motivate $\mathcal{K}$ still persists. In other words, simply
assuming that $\mathcal{K}$ is only an integration constant, as obtained integrating the continuity
equation, is not enough to physically motivate the dark fluid existence. Hence, the problem belongs
to understanding a fundamental theory for the dark fluid which is up to now not completely known, although
reasonable advances have been already presented in the literature (see for example \cite{Liddle:2006qz,Reyes:2011ij}).
This fundamental theory should motivate $\mathcal{K}$ and the dark fluid itself. In
Sec.~\ref{Section:BackCosm}, we provide thermodynamical interpretations of $\mathcal{K}$,
able to physically clarify its role in the framework of cosmology and so we propose a self-consistent approach to characterize the dark fluid.


\section{Thermodynamical systems with constant pressure}  \label{Section:general}

We  now describe the thermodynamical consequences of systems with constant pressure.
Hence, we address a very general thermodynamical scenario, whereas later on, we will use
intensive variables to correctly describe the dark fluid phenomenology in
the framework of cosmology. In other words, to better understand which
consequences occur in the framework of dark fluid's approach, we first
discuss some general features of thermodynamic systems provided with constant
pressure. For such systems, the mechanical EoS is restricted to be
\begin{equation} \label{General:TEOS}
 P=P_0,
\end{equation}
where $P$ is the pressure and $P_0$ is a constant. Besides the
above hypothesis, we regard the generic thermodynamic system by means of
the following extensive variables: the internal (or {\it first-law})
energy $U$, from now on simply called the \emph{energy}, the volume $V$ and the
number of constituents, or {\it particles}, $N$. Hereafter, we do not
distinguish between particles and constituents. We are interested in
finding the general properties of this system when the EoS for pressure is restricted to Eq. (\ref{General:TEOS}). A thermodynamical
system in equilibrium can be completely specified by a fundamental
equation, thus we look for a thermodynamical potential from which
Eq.~(\ref{General:TEOS}) may be derived.
To this end we work in the entropy representation, the Gibbs equation reads
\begin{equation} \label{General:GibbsRepS}
dS = \frac{1}{T}dU + \frac{P}{T}dV -\frac{\mu}{T}dN\,,
\end{equation}
where $S(U,V,N)$ is a function of the extensive variables. On the other hand we can
write\footnote{In this section,
we will clump the notation by explicitly showing the dependence of every
considered function.}
\begin{equation}\label{General:dS}
dS =\left( \frac{\partial S}{\partial U} \right)_{V,N} dU+
\left(  \frac{\partial S}{\partial V} \right)_{U,N} dV +\left(
\frac{\partial S}{\partial N} \right)_{U,V} dN\,.
\end{equation}
From Eqs.~(\ref{General:GibbsRepS}) and (\ref{General:dS}) it follows that a necessary and
sufficient condition for guaranteeing the pressure to be constant is
\begin{equation} \label{General:ConstPCondition}
\frac{\partial S}{\partial V} - P_0  \frac{\partial S }{\partial U} = 0\,.
\end{equation}
Thus, the entropy turns out to be $S(U,V,N) = k_B f(U + P_0 V,N)$. Here,
the Boltzmann constant $k_B$ is explicitly reported to have the function $f$ dimensionless.
Demanding a homogeneous of first degree entropy function,
that is, $S(\lambda U,\lambda V,\lambda N) = \lambda S(U,V,N)$, and setting
$\lambda = N^{-1}$, we obtain
\begin{equation} \label{General:Entropy}
S(U,V,N) = N k_B f \left( \frac{U+P_0 V}{N} \right)\,.
\end{equation}
At this stage, it is helpful to set
\begin{equation}
 x \equiv \frac{U + P_0 V}{N}, \qquad F(x) \equiv \frac{df}{dx}\,.
\end{equation}
The entropy is therefore drawn up as $S=N k_B f(x)$ and from
Eqs.~(\ref{General:GibbsRepS}) and (\ref{General:dS}),
we obtain the corresponding equations of state
\begin{equation}
 \frac{1}{T} = \left(\frac{\partial S}{\partial U} \right)_{V,N} = k_B
F(x)\,, \label{general:1overT}
\end{equation}
\begin{equation}
 \frac{P}{T} = \left(\frac{\partial S}{\partial V} \right)_{U,N} = k_B P_0
F(x)\,,  \label{general:PoverT}
\end{equation}
\begin{equation}
 \frac{\mu}{T} = -\left(\frac{\partial S}{\partial N} \right)_{U,V}
 = -k_B \big(f(x) - x F(x) \big)\,. \label{general:MuoverT}
\end{equation}
The case $P=P_0$ follows from Eqs.~(\ref{general:1overT}) and
(\ref{general:PoverT})
as expected, whilst the chemical EoS (\ref{general:MuoverT})
has the form of a Legendre transformation, and as a consequence the
Gibbs-Duhem equation, given by
\begin{equation}
d \left( \frac{\mu}{T}\right) = k_B x  F'(x) dx\,,
\end{equation}
does not furnish additional thermodynamic information. From
Eq.~(\ref{general:1overT}), we find that $x$ can be written as a function
of the temperature only
\begin{equation} \label{xofT}
 x = F^{-1}\left( \frac{1}{k_B T}\right)\,,
\end{equation}
where $F^{-1}$ is the inverse function of $F(x)$. The entropy, written as
a function of the temperature $T$, becomes
\begin{equation}
 S(P,T,N) = N k_B f \left( F^{-1} \left( \frac{1}{k_B T} \right)
\right)\,. \label{general:SPTN}
\end{equation}
For a fixed number of constituents $N$, the entropy becomes a function of the
temperature only. In this case, the forthcoming result is that for
equilibrium thermodynamical systems with constant pressure isothermal and
adiabatic processes coincide.

Moreover, the \emph{stability conditions}
\begin{equation} \label{conditions89}
 \partial^2_{UU}S \leq 0\,, \qquad \partial^2_{VV} S \leq 0, \qquad
\text{and} \qquad \det S_{,ij} \geq 0\,,
\end{equation}
are satisfied for $F'(x) < 0$.

Cumbersome algebra provides  $U(S,V,N) = N f^{-1} ( S/k_B N) - P_0 V$ and
$H(S,P,N) = N f^{-1} ( S/k_B N)$,
where $H$ is the enthalpy, and by using Eq.~(\ref{general:SPTN}),
\begin{equation}
 U(T,V,N) = N F^{-1} \left( \frac{1}{k_B T}\right) - P_0V\,,
\label{general:UTVN}
\end{equation}
and
\begin{equation}
 H(T,P,N) = N F^{-1} \left( \frac{1}{k_B T}\right) \equiv N h(T)\,.
\label{general:HTPN}
\end{equation}
In the last equality we defined the enthalpy per particle $h(T)$.
From Eqs.~(\ref{general:UTVN}) and (\ref{general:HTPN}), we find out the
heat capacities
\begin{equation}   \label{general:CV}
 C_V \equiv \left(\frac{\partial U}{\partial T} \right)_{V,N} =
-\frac{N}{k_B T^2} \frac{1}{F'(F^{-1}(1/k_B T))}\,,
\end{equation}
and
\begin{equation}  \label{general:CP}
 C_P \equiv \left(\frac{\partial H}{\partial T} \right)_{P,N} = C_V\,.
\end{equation}
From Eq.~(\ref{general:CV}), one infers that $C_V \geq 0$ in case
$F'(x) < 0$. Thus, in addition to the condition that entropy
is a monotonic growing-function of $U$, and therefore of $x$, if the
criterion of stability is appealed,
$f(x)$ is naturally convex, i.e. $F(x) > 0$ and $F'(x) < 0$.

A relevant outcome we got is that both heat capacities coincide,
whereas the adiabatic index is $\gamma \equiv C_P/C_V= 1$, as
expected since the pressure is constant.

By using Eq.~(\ref{general:UTVN}), we obtain for the energy density $\rho \equiv U/V$
\begin{equation}
 \rho = -P_0 + \frac{N h(T)}{V}\,.  \label{general:rho1}
\end{equation}
Above we demonstrated that, albeit we involved an adiabatic expansion, the
temperature
is a constant. It follows that for a positive energy density, for
arbitrary ratios $V/N$, the condition
\begin{equation}
 P_0 < 0\,,
  \label{general:PRestricted}
\end{equation}
must hold.
The pressure of the system is a negative constant quantity as a
consequence of the thermodynamics here presented.

The case of negative pressure is not unusual in nature. Examples of systems in
which this condition holds are given in \cite{Wheeler:Nature,Coupin:bib,Stanley:2002281} and references therein. In those
examples, the particular choices of thermodynamic
conditions naturally portray the fact that the pressure is negative.

Another example is derived from Geometrothermodynamics
\cite{Quevedo:2006xk}. This formalism
has been proposed to describe thermodynamical system on pure geometrical
grounds. Geometrothermodynamics also provides a way to obtain fundamental equations
as solutions of an extremal surfaces principle \cite{Vazquez:2011ia}. The cosmological consequences of one of
such solutions have been studied in  \cite{Aviles:2012et}, leading to an extension of
the generalized Chaplygin gas, which for some specific choice of their free parameters
reduces to a dark fluid with entropy

\begin{equation} \label{GTD:Entropy}
 S = S_0 N \ln \left( \frac{U + cV}{N} \right),
\end{equation}
which is of the form of Eq.~(\ref{General:Entropy}), and consequently the
pressure becomes constant.
Using the above formalism, we obtain the entropy of the
system as a function of the temperature $S(T,N) = S_0 N \ln (S_0 T)$.
Note that this entropy results unbounded from below, thus the Nerst
postulate is violated in this particular example.
The resulting energy density is
\begin{equation}\label{GTD:rhoTV}
 \rho = -P_0 + \frac{N S_0 T}{V}\,,
\end{equation}
which, if compared to Eq. (\ref{general:rho1}), provides $h(T)=S_0T$.

\subsection{The emergent cosmological constant}

In the case of zero constant pressure, the system described by
Eq.~(\ref{general:rho1}),
gives as density a direct functional dependence in terms of the volume and
temperature as:
\begin{equation}\label{ZP:rhoTV}
 \rho =  \frac{N h(T)}{V}.
\end{equation}
One recognizes that the former represents the case of dust particles if
the system is maintained at constant temperature.
Here, the dark fluid reduces to a pure dark matter component. This would
be also the case of non-interacting baryonic
particles.

Obtaining a pure dark energy term is subtler. It corresponds to the
particular case in which the enthalpy vanishes, {\it i.e.}
\begin{equation} \label{vacc:hofT}
h(T) = F^{-1}\left( \frac{1}{k_B T} \right) = 0\,.
\end{equation}
Generally, the above equality is valid at one specific temperature,
hereafter $T_\Lambda$. Inverting the equation to get $T_\Lambda$, one gets
\begin{equation}
 T_\Lambda = \frac{1}{k_B F(x)|_{x=0}}\,,
\end{equation}
implying at $x=0$,  $U=-P_0 V$, or equivalently
\begin{equation}
 \rho = -P_0\,.
\end{equation}
Immediately, we can notice that if the temperature takes the constant
value $T_\Lambda$, any  adiabatic processes maintain the same temperature
at all stages
of the universe evolution, providing an overall both adiabatic and
isothermal process. However, it is clear that any deviation from
equilibrium increases the entropy. Due to this fact a corresponding
temperature change is expected. In this situation, Eq.~(\ref{vacc:hofT})
is no longer satisfied, and we therefore refer to pure dark energy as a
{\it unstable fine-tuning} case.


\section{Intensive variables and background Cosmology}
\label{Section:BackCosm}

A complete and general picture accounting thermodynamics in curved
space-time has not definitively forecasted.
Before proposing an approach able to describe such a situation, it is
convenient to reobtain all results of
Sec.~\ref{Section:general} by means of intensive variables, which show the
advantage to be locally defined.
We thus define the specific entropy (per particle), $\sigma \equiv S/N$,
and the particle density $n \equiv N/V$. The entropy therefore reads
\begin{equation}
\sigma = k_B f(x)\,,
\end{equation}
and its variation in terms of first order differential form
\begin{equation}\label{kjhfkhdfkjdh}
d\sigma = k_B F(x) dx\,,
\end{equation}
where now $x= (\rho + P_0)n^{-1}$. Using Eqs.~(\ref{xofT}) and (\ref{general:CV}) above, it follows:
\begin{equation} \label{IV:dsigma}
 d\sigma = \frac{c_V(T)}{T}dT\,,
\end{equation}
where $c_V(T) \equiv C_V(T,N) / N $ is the specific heat capacity (per
particle) at constant volume, with its
definition given by Eq.~(\ref{general:CV}) ---Analogous definition for
$c_P(T)$ will be used below.
In this case we derived from local variables the result that the entropy
is a function of the temperature only.
The corresponding Gibbs equation (\ref{General:GibbsRepS}) becomes

\begin{equation}
 T d\sigma = \frac{1}{n} \left[ d\rho - \frac{ \rho + P}{n} dn \right]\,,
\label{IV:Tdsigma}
\end{equation}
or, altenatively,
\begin{equation}
 T d\sigma  =  \frac{1}{n}\left(\frac{\partial \rho}{\partial T} \right)_n
dT+ \frac{1}{n} \left[ \left(\frac{\partial \rho}{\partial n} \right)_T
-  \frac{\rho + P}{n}   \right]dn\,.     \label{IV:Tdsigma2}
\end{equation}
By comparing Eqs.~(\ref{IV:dsigma}) and (\ref{IV:Tdsigma2}) we obtain
\begin{equation} \label{rhoofTandn}
 \rho = -P_0 + h(T)n\,,
\end{equation}
which confirms Eq.~(\ref{general:rho1}). Consequences on background
cosmology are easily accounted by appealing
local thermodynamical equilibrium.\footnote{The local equilibrium is not
strictly necessary if the dark fluid is the only energy component. A
necessary and sufficient condition for a corresponding thermodynamical
description is that the differential
form $Td\sigma$ is integrable. In flat space, it is easily satisfied (and for
some authors the temperature is defined as an integration factor),
but in curved space-times, it is not necessarily true,
as studied in \cite{Krasinski:1997zz,Hernandez:2007jb}. Nevertheless, for the dark fluid
Eq.~(\ref{IV:Tdsigma}) can be rewritten as
$Td\sigma = d(\rho/n) + P d(1/n) = d((\rho + P_0)/n)$ and therefore it is integrable.
However, we do not follow this approach, since we will consider
dissipative processes in Sec.~\ref{Sect:EntProd}.}
To do so, let us divide
the whole system at fixed FRW-coordinate time $t$ in spatial cell
subsystems each with comoving volume $V_0$
and constituent number $N$, wherein local equilibrium holds.
In this picture, the processes are considered quasi-static and the
thermodynamic equilibrium is recovered at every stage of local evolution,
although a whole equilibrium is not necessarily addressed.
The corresponding physical volume $V$ is related to the scale factor,
becoming $V(a) = V_0 (a/a_0)^3$, and then we obtain, for the energy
density evolution
\begin{equation} \label{BackCosm:energydensity}
\rho(a) = \frac{\rho_0}{1+\mathcal{K}_0} \left(1 +
\frac{\mathcal{K}(T,N)}{a^3} \right)\,.
\end{equation}
where $\mathcal{K}$ is a function of $T$ and $N$
\begin{equation}  \label{BackCosm:KTN}
 \mathcal{K}(T,N) = -\frac{1}{P_0} N h(T) \frac{a^3_0}{V_0} \equiv - v^{-1} \frac{h(T)}{P_0} \,,
\end{equation}
Notice in the last equality we defined the specific comoving volume per constituent, $v \equiv V_0/a_0^3 N$,
and $\mathcal{K}_0 \equiv \mathcal{K}(a_0)$.
Soon, it is evident that Eq. (\ref{BackCosm:energydensity}) generalizes
Eq.~(\ref{DarkFluidRev:rhodEv}),
in which non-adiabatic contributions have been included through the term
$\mathcal{K}(T,N)$.

In case of constant $N$, we choose $V_0$ such that $v =1$ and we absorb the volume units in the definition of $h(T)$,
simplifying our notation.
Further, in the absence of bulk viscosity, the universe expansion history
follows an adiabatic local
equilibrium process, otherwise the symmetries of homogeneity and isotropy
would break, as we will
demonstrate in Sec.~\ref{Sect:EntProd}. We showed in
Sec.~\ref{Section:general} that adiabatic processes
are also isothermal for a system with constant pressure. Thus, the
enthalpy per
particle becomes a constant as well,  $h(T) = h(T_0)$, and
\begin{equation}
 \mathcal{K} = -\frac{h(T_0)}{P_0}\,.
\end{equation}
Hence, for an adiabatic process, with a fixed number of constituents $N$,
$\mathcal{K}$ is a constant equal to the enthalpy over the pressure,
contained in a unitary comoving volume, recovering
Eq.~(\ref{DarkFluidRev:rhodEv}).
A relevant consequence of our formalism is that $\mathcal{K}$ is no
longer an integration constant, since
it was derived from thermodynamical first principles. This may be reviewed
as a first explanation of its physical meaning, permitting to better
justify the dark fluid existence.

Using Eqs.~(\ref{DarkFluidRev:DFLCDMId}), a comparison with the
$\Lambda$CDM model gives
\begin{eqnarray}
 h(T_0) &=& \rho_c \Omega_{\text{dm}}, \label{Eqa1}\\
 P_0 &=& \rho_c \Omega_{\Lambda},  \label{Eqa2}
\end{eqnarray}
In the absence of a thermodynamical fundamental equation, we cannot invert
Eq.~(\ref{Eqa1}) to obtain the value of $T_0$. However, we could appeal
to several
possible considerations on how to treat both the dark fluid and
``visible''  sector. The first possibility is to assume that the dark
fluid and the ``visible''  sector had a common thermal origin. Afterwards,
they decoupled at a precise energy scale,  $k_B T_{*}$, thereafter the
temperature of the dark fluid was maintained constant by an adiabatic
expansion and today its value is $T_0 = T_{*}$. In this scenario, the
decoupling era had to be at very early times, so that
$a- a_{*} \simeq a$, where $a_{*}$ is the
decoupling scale factor and $a$ lies on the range considered throughout this article.
A second approach is that no interactions between
the dark and ``visible'' sectors were present at all. Thus, $T_{*}$
becomes the temperature at which our description starts to be valid, whereas $a_{*}$
the corresponding scale factor. Moreover, since the $\Lambda$CDM has been
probed to match data, especially at early times by cosmic microwave
background observations, we also conclude that $a_{*} \rightarrow 0$. Both
approaches provide that the time $t_{*}$, defined by $a_{*} =
a(t_{*})$, is very small. The only difference is that, in the first
scenario the temperature $T_{*}$ is very large and is considered as a {\it
freezing} temperature $T_{*} > \mathcal{O}(\text{MeV})/k_B$ (above the primordial nucleosynthesis scale), while
in the second case its value is not constrained by these considerations.

Afterwards, we extend the framework of constant pressure, enabling it to
vary.
In an adiabatic process, it can be shown that the following relation holds
\cite{Weinberg:1971mx}:
\begin{equation} \label{Deriv:dTdt}
 \left( \frac{\partial \rho}{\partial T}\right)_n \frac{\dot{T}}{T} =
\left( \frac{\partial P}{\partial T}\right)_n \frac{\dot{n}}{n}\,.
\end{equation}
In the case of constant pressure, the right hand side of Eq.
(\ref{Deriv:dTdt}) vanishes and we obtain again that the temperature is
a constant. We note that the same happens for a baryonic dust component and
the well-known relation, $T_{b} \propto a^{-2}$, is naturally recovered
if we consider the kinetic theory of non-relativistic
particles\footnote{This result can also be derived from thermodynamical
arguments, where
$P_{b}=nkT$ and $\rho_{b} = n ( m + 3kT/2)$ are good approximations when
$k_B T\ll m$, then it follows from Eq.~(\ref{Deriv:dTdt})
that $T_{b} \propto a^{-2}$.}. The artificial EoS, $P_{dust} = 0$,
although a good approximation, led us to a non-sense conclusion providing
a constant dust temperature. We can reach analogous results for the case of constant pressure. If we
naively extrapolate
the above ideas to the case of the dark fluid, we may
simply set $P=P_0 + nkT$, and $\rho = n( m + 3kT/2)$, obtaining again $T
\propto a^{-2}$. However, this approach lies on kinetic theory considerations, which may not apply
for the dark fluid case.


\section{Entropy production} \label{Sect:EntProd}

In this section, we investigate out-of-equilibrium processes enabling the dark fluid to deviate from its standard evolution,
which degenerates with the $\Lambda$CDM model. First, since a complete description of the dark fluid is so far unknown, we develop an
alternative approach and we summarize some interesting points, which will
be helpful when we discuss entropy production generated by out-of-equilibrium processes. Let us assume that the constant pressure shifts to
$P_* = P_0 + \pi$.
The effect of the {\it dynamical}
pressure $P_*$ is to induce a temperature change given by
\begin{equation} \label{Deriv:dTdtPi}
\frac{\dot{T}}{T} = \left( \frac{\partial \pi}{\partial \rho}\right)_n
\frac{\dot{n}}{n}\,,
\end{equation}
obtaining that the temperature is no longer a constant. The term $\pi$ is named bulk viscosity, in the context of dissipative processes of
irreversible thermodynamics. In such a case, Eq.~(\ref{Deriv:dTdt}) is no longer valid
because of the corresponding entropy production. In
Sec.~\ref{Sect:EntProd}, we will see that Eq.~(\ref{Deriv:dTdtPi}) holds
only for the case in which $\pi(T,n)$ is a linear function of $T$.
In this situation, we may assume $\pi = \pi_0(n) + \pi_1 (n) T$, and with
the aid of Eq.~(\ref{general:CP}), we find that Eq.~(\ref{Deriv:dTdtPi})
becomes
\begin{equation}
  \frac{ c_P(T) dT}{T} =  \frac{\pi_1 (n) \, dn }{n^2}\,.
\end{equation}
In relativistic Eckart's theory \cite{Eckart:19403}, $\pi_1$ is negative, and
assuming the number density decays with universe's expansion, we
obtain the well-known result that viscosity is capable of raising the
fluid temperature. To properly introduce the out-of-equilibrium processes,
it is convenient to split the stress-energy tensor in a piece that
correspond to equilibrium states plus a piece that contains the
dissipative terms as
\begin{equation} \label{EntProd:SETensor}
 T^{\mu\nu} = T^{\mu\nu}_{equi} + D^{\mu\nu}\,.
\end{equation}
It is always possible to choose a four vector velocity field $u^\mu$ for
which the energy density $\rho$ and
the particle density number $n$ coincide with their
equilibrium values \cite{Eckart:19403,Weinberg:1971mx}, dealing with the commonly named {\it N-frame}. We decompose
$T^{\mu\nu}$ in proper components of $u^\mu$, giving
\begin{equation}
 T^{\mu\nu}_{equi} = \rho u^\mu u^{\nu} + P h^{\mu\nu}\,,
\end{equation}
\begin{equation}
 D^{\mu\nu} = \pi h^{\mu\nu} + q^{\mu}u^{\nu} + q^{\nu}u^{\mu} +
\pi^{\mu\nu}\,,
\end{equation}
where
\begin{equation}
 h_{\mu\nu} = g_{\mu\nu} + u_{\mu}u_{\nu}\,,
\end{equation}
is the projector tensor on spatial hypersurfaces normal to $u^\mu$ and
\begin{equation}
\pi = \frac{1}{3} h_{\mu\nu} D^{\mu\nu}\,,
\end{equation}
\begin{equation}
\pi^{\mu\nu} = D^{\langle \mu\nu \rangle} \equiv
 \frac{1}{2}h^{\mu}_{\,\,\alpha}h^{\nu}_{\,\,\beta}\left( D^{\alpha\beta}
+ D^{\beta\alpha}\right)
 - \frac{1}{3} h_{\alpha\beta} D^{\alpha\beta}h^{\mu\nu}\,,
\end{equation}
\begin{equation}
 q^{\nu} = - u^{\alpha} h^{\beta\nu}D_{\alpha\beta}\,,
\end{equation}
are the dissipative terms, which can be identified with bulk viscosity,
shear viscosity and heat flow, respectively.
Thus,
\begin{equation}
 q^{\nu}u_\nu = \pi^{\mu\nu}u_\nu = \pi^\alpha_{\,\,\alpha}= 0, \qquad
h^{\nu}_{\,\,\alpha} q^\alpha = q^{\nu}, \qquad
 h^{\mu}_{\,\,\alpha} \pi^{\alpha\nu} = \pi^{\mu\nu}, \qquad  \pi^{\mu\nu}
=  \pi^{\langle \mu\nu \rangle}\,.
\end{equation}
From the conservation equations $\nabla_\mu T^{\mu\alpha} = 0$, we obtain
by contracting with $u_{\alpha}$, the continuity equation
\begin{equation} \label{EntProd:ContEq}
 \dot{\rho} + \Theta \big( \rho + P + \pi \big) + \big( \nabla_\mu +
\dot{u}_\mu \big) q^\mu
 + \pi^{\mu\nu}\sigma_{\mu\nu}=0\,,
\end{equation}
and by contracting with $h^{\nu}_{\,\,\alpha}$ the three hydrodynamical
Euler equations
\begin{equation} \label{EntProd:HydrodEulerEq}
 h^{\mu\nu}\nabla_{\mu} \big( P + \pi \big) + \big( \rho + P + \pi
\big)\dot{u}^{\nu}
 + h^{\nu}_{\,\,\alpha} \dot{q}^\alpha + \left( \frac{4}{3} \Theta
h^{\nu}_{\,\,\alpha} + \sigma^{\nu}_{\,\,\alpha}
 + \omega^{\nu}_{\,\,\alpha} \right) q^\alpha + h^{\nu}_{\,\,\alpha}
\nabla_{\mu} \pi^{\mu\alpha}= 0\,.
\end{equation}
The dot over an arbitrary tensor is the covariant derivative along the
integral curves of
$u^{\mu}$ ($\dot{\mathbf{A}} \equiv u^{\mu}\nabla_{\mu}\mathbf{A}$);
additionally $\Theta = \nabla_{\mu}u^{\mu}$,
$\sigma_{\mu\nu} = \nabla_{\langle \nu} u_{\mu\rangle}$,
$\omega_{\alpha\beta} = h^{\mu}_{\,\,\alpha}h^{\nu}_{\,\,\beta}
\nabla_{[\nu}u_{\mu]}$ are the expansion scalar, the shear and
the vorticity of the world lines defined by $u^{\mu}$, respectively.

In this {\it N-frame}, the particle density flux is
\begin{equation} \label{EntProd:Defnmu}
 n^\mu = nu^{\mu}\,,
\end{equation}
therefore, the conservation of particle density flux $\nabla_\mu n^{\mu} =0$ gives
\begin{equation} \label{EntProd:ConsN}
 \dot{n} + \Theta n = 0\,.
\end{equation}
In local equilibrium, the entropy flux is given by
\begin{equation} \label{EntProd:EntropyFluxEquil}
 S^{\mu}_{equil} = n \sigma u^{\mu}\,,
\end{equation}
thus, the entropy density is given by $-u_{\mu}S^{\mu}_{equil} = n \sigma$.
Further, out-of-equilibrium, the entropy flux vector becomes
\begin{equation} \label{EntProd:EntropyFlux}
 S^{\mu} = n \sigma u^{\mu} + \frac{q^\mu}{T} -
Q^\mu(\pi,q^\alpha,\pi^{\alpha\beta};u^{\alpha})\,,
\end{equation}
where $Q^{\mu}$ is the most general function quadratic in the dissipative terms and $\sigma$ is
the equilibrium specific entropy. The Eckart theory
lies on neglecting $Q^\mu$, although this leads to instantaneous propagation of heat \cite{Israel:1976tn} and undesired
instabilities in the system \cite{Hiscock:1983zz}. The introduction of $Q^\mu$ is
necessary to account for those effects
apart from the case where gradients of $\pi$, $q^\alpha$ and
$\pi^{\mu\nu}$ are negligible \cite{Maartens:1996vi}. By the Gibbs equation (\ref{IV:Tdsigma})
and  the continuity equation (\ref{EntProd:ContEq}), the
rate of change of the entropy per particle is
\begin{equation} \label{EntProd:ratesigma}
 n T \dot{\sigma} = -\pi \Theta - \big( \nabla_\mu + \dot{u}_\mu \big) q^\mu
 - \pi^{\mu\nu}\sigma_{\mu\nu}\,,
\end{equation}
and the entropy flux becomes
\begin{equation} \label{EntProd:EntropyFlux2}
 T\nabla_\mu S^\mu = - \pi \Theta - \left( \frac{\partial_\mu T}{T}
 + \dot{u}_\mu \right) q^\mu - \pi^{\mu\nu}\sigma_{\mu\nu} - T \nabla_\mu
Q^\mu\,.
\end{equation}
In relativistic thermodynamics the second law of thermodynamics is written
as an inequality on the divergence
of the entropy flux \cite{Eckart:19403,Weinberg:1971mx,Misner:1974qy}
\begin{equation} \label{EntProd:2law}
 \nabla_{\mu}S^\mu \geq 0\,,
\end{equation}
that is, the rate at which the entropy is generated can either vanish or be positive \cite{Misner:1974qy}.

\subsection{The temperature rate of change}

\noindent Now, let us consider the rate of change along integral curves of $u^{\mu}$
of the energy density $\rho(n,T)$,
\begin{equation}
 \dot{\rho} = \left( \frac{\partial \rho}{\partial n} \right)_T \dot{n} +
\left( \frac{\partial \rho}{\partial T} \right)_n \dot{T}.
\end{equation}
Using the identity
\begin{equation}
 \left(\frac{\partial P}{\partial T}\right)_n \dot{T} = \rho + P - n
\left(\frac{\partial \rho}{\partial T}\right)_n,
\end{equation}
and the Gibbs equation, we obtain for the rate change of the temperature
\begin{equation} \label{EntProd:rateT1}
 \left(\frac{\partial \rho}{\partial T}\right)_n \frac{\dot{T}}{T} =
n\dot{\sigma} +
 \left(\frac{\partial P}{\partial T}\right)_n \frac{\dot{n}}{n}.
\end{equation}
Using now Eq.~(\ref{EntProd:ratesigma})
\begin{equation} \label{EntProd:rateT2}
 \left(\frac{\partial \rho}{\partial T}\right)_n \dot{T} = -\pi \Theta -
\big( \nabla_\mu + \dot{u}_\mu \big) q^\mu
 - \pi^{\mu\nu}\sigma_{\mu\nu}
+ T \left(\frac{\partial P}{\partial T}\right)_n \frac{\dot{n}}{n}.
\end{equation}
So far, the analysis performed in Eq.~(\ref{EntProd:rateT2}) is general, and hence, applies for any fluid described by
the stress-energy tensor (\ref{EntProd:SETensor}). Further, one can notice that
Eq.~(\ref{Deriv:dTdtPi}) is recovered when $\pi$
is a linear function of the temperature.
Therefore, the effect of bulk viscosity is not to shift
the pressure from $P$ to $P + \pi$ only, but also it
affects the temperature rate change differently from the pressure, as showed in Eq.~(\ref{EntProd:rateT2}). Applying this result to the case of constant pressure, and having
$(\partial \rho/\partial T)_n = n c_v(T) = n c_p(T)$,
Eq.~(\ref{EntProd:rateT2}) then reduces to
\begin{equation} \label{EntProd:rateT3}
 n c_p(T) \dot{T} =  -\pi \Theta - \big( \nabla_\mu + \dot{u}_\mu \big) q^\mu
 - \pi^{\mu\nu}\sigma_{\mu\nu}.
\end{equation}

\subsection{The case of background Cosmology and phantom behavior}

\noindent Now, let us consider the effect of the dissipative terms in a spatial
homogeneous and isotropic  universe. We write
the FRW spatially flat metric as
\begin{equation}
 ds^2 = -dt^2 + a^2(t)\delta_{ij}dx^idx^j\,.
\end{equation}
The time coordinate $t$ coincides with the proper time of the
free fall observers comoving with the Hubble flow.
In these coordinates, the four velocity is written as $u^{\mu} = (1,{\bf 0})$, and a dot over a scalar quantity becomes the partial derivative
with respect to $t$, e.g. for a scalar function $f$, $\dot{f} = \partial_t f$.

The only dissipative term, compatible with this geometry is the bulk
viscosity, thus Eq.~(\ref{EntProd:ratesigma}) becomes
\begin{equation}
 \dot{\sigma} = -\frac{\pi \Theta}{nT}\,,
\end{equation}
and the Gibbs equation leads to
\begin{equation}
 nT\dot{\sigma} =  - \pi \Theta = \dot{\rho} + \Theta (\rho + P)\,.
\end{equation}
Since we are considering a FRW metric the expansion scalar is $\Theta = 3
H = 3 \dot{a}/a$, and appealing conservation
of particle flux (see Eq.~(\ref{EntProd:ConsN}) above) we obtain
$n = n_0 a^3_0/a^3$. By Eq.~(\ref{EntProd:rateT3}) the temperature change is
\begin{equation} \label{EntProd:rateTFRW}
  c_p(T) \dot{T} =  -\frac{\pi \Theta }{n } = - 3a^2 \dot{a} \pi\,.
\end{equation}
Soon, we notice that the second law of thermodynamics requires
\begin{equation} \label{EntProd:2lawFRW}
 \nabla_\mu S^{\mu} = n \dot{\sigma} - \nabla_\mu Q^\mu(\pi^2;u^{\alpha})
\ge 0\,.
\end{equation}
In Eckart's theory, the quadratic term $Q^\mu$ is neglected and customarily one can
choose the algebraical equation
\begin{equation} \label{BVET}
\pi = -3H \zeta\,,
\end{equation}
with $\zeta>0$, which reduces to the Newtonian Stokes law $\pi = -\zeta
\vec{\nabla}\cdot\vec{v}$ in the non relativistic limit, and ensures the
positivity of the entropy flux divergence. Thereafter,
$\zeta = \zeta(\rho) = \rho^{-1/2} g(\rho)$, thus if non-dark components
are neglected, $\rho^{-1/2}H$ is a constant and $\pi$ becomes a function of
the energy density only. This is equivalent to simply choose $\pi =
\pi(a)$ in a manner that the condition (\ref{EntProd:2lawFRW})
is satisfied. Thus, anagously, in Eckart's theory, we could simply set $\pi =
\pi(a)<0$ as well.\footnote{
The correspondence between this approach and Eckart's theory can be seen as follows:
Assume the solution $\rho = \rho(a)$ has an inverse $a=a(\rho)$ (if not, consider the inverse piecewise). From
the continuity equation (\ref{ContEqwV}) we obtain
\begin{displaymath}
 \pi(\rho) = -\frac{1}{3} \left( \frac{d \ln a(\rho) }{d\rho} \right)^{-1} - \rho - P_0\,.
\end{displaymath}
In Eckart's theory $\pi(\rho,H) = -3 H \zeta(\rho)$,
with the aid of Friedmann equation in the case in which other fluids than the dark fluid can be neglected we obtain
\begin{displaymath}
 \zeta(\rho) = \sqrt{\frac{3}{8\pi G}} \rho^{-1/2} \left(\frac{1}{9} \left( \frac{d \ln a(\rho) }{d\rho} \right)^{-1} + \frac{1}{3}(\rho + P_0) \right)\,.
\end{displaymath}
In the appendix of this article,
we work out an exact {\it Big Rip} solution in Eckart's theory and we explicitly show the correspondence to our formalism.
}
However, if  early time evolution is considered,
we cannot neglect the ``visible'' sources and $\rho^{-1/2}H$ is no longer
constant. A particular remark is that several authors prefer to give a dependence $\zeta=
\zeta(\rho_{total})$, where $\rho_{total}$ is the total energy density,
implying interactions among the dark and ``visible'' sectors. In so doing,
the dependence of $\pi$ on $H$ can be avoided.

\noindent In the following, we adopt the general approach in which the bulk
viscosity $\pi$ is a function of $\rho$ and $H$, or a function
of $a$ and $\dot{a}$,
\begin{equation} \label{piaadot}
 \pi = \pi(a,\dot{a})\,.
\end{equation}
In MIS causal theory, $Q^\mu$ is not neglected and being
quadratic in the bulk viscosity one arrives to
\begin{equation} \label{ISTEF}
 \nabla_\mu S^\mu = -\frac{\pi}{T} \left[ 3 H + \beta_0 \dot{\pi} +
\frac{1}{2}T \nabla_\alpha \left( \frac{\beta_0}{T} u^{\alpha} \right)
\pi \right]\,,
\end{equation}
where $\beta_0$ is a phenomenological non-negative coefficient ---it is
the coefficient of the $\pi^2$ term in $Q^\mu$.
The simplest way to satisfy the second law of thermodynamics is to impose
\begin{equation} \label{ISTTE}
 \tau_0 \dot{\pi} + \pi = - 3 \zeta H - \left[ \frac{1}{2} \zeta  T
\nabla_\alpha  \left( \frac{\beta_0}{T} u^{\alpha} \right) \pi \right]\,,
\end{equation}
which is obtained by equating the terms inside the brackets of
Eq.~(\ref{ISTEF}) to $-\pi$, and the $\zeta$ and $\tau_0$ coefficients are
related among them by  $\beta_0 = \tau_0/\zeta$, having the meaning of
bulk viscosity strength and relaxation time, respectively.
This first order differential equation is well supported by arguments derived from the standard kinetic theory
\cite{Israel1976213}, which does not necessarily have to apply to the dark
fluid.

In kinetic theory $\tau_0$ is identified with the mean free path of the
particles, and if $\tau_0 \ll H$, Eq.~(\ref{ISTTE}) reduces to \cite{Maartens:1996vi}
\begin{equation} \label{ISTTET}
 \tau_0 \dot{\pi} + \pi = - 3 \zeta H\,.
\end{equation}
This equation is the transport equation in the {\it truncated}
MIS theory for bulk viscosity. Given the two phenomenological
functions
$\zeta$ and $\tau_0$ we can solve for $\pi$ and $\rho$. Alternatively
we can adopt a phenomenological $ \pi =  \pi(a,\dot{a})$ and
Eq.~(\ref{ISTTE}) (or Eq.~(\ref{ISTTET})) becomes a constriction on
$\zeta$ and $\tau_0$.

\noindent Now, giving $dh = c_p(T) dT$ for constant pressure systems, we obtain from
Eq.~(\ref{EntProd:rateTFRW}) the enthalpy change
\begin{equation} \label{doth}
 \dot{h} = - 3 a^2 \pi(a,\dot{a}) \dot{a}\,.
\end{equation}
Our main goal is to find out a general solution $h=h(a)$. Unfortunately, this is impossible for the general
case of Eq.~(\ref{piaadot}) by using only Eq.~(\ref{doth}), we thus have
to integrate the complete system of equations, that is, the Friedmann
equation
\begin{equation}
 H^2 = \frac{8\pi G}{3} \rho_{total}\,,
\end{equation}
the continuity equation (\ref{EntProd:ContEq}), which we appropriately rewrite as
\begin{equation} \label{ContEqwV}
 \dot{\rho_d} + 3 H(\rho_d + P_0 + \pi ) = 0\,,
\end{equation}
and the corresponding continuity equations for the non-dark components. To
close the system we need the equation for the bulk viscosity,
which is generally chosen to be one of (\ref{BVET}), (\ref{ISTTE}) or
(\ref{ISTTET}) for Eckart, MIS or truncated MIS,
respectively.
Nevertheless, we emphasize that those restrictions, although well
motivated for standard non-dark fluids, are not strictly necessarily, they are just the easiest way
to ensure the positivity of the entropy flux divergence.

In the special case where $\pi$ is a function of the scale factor only,
$\pi = \pi(a)$,
Eq.~(\ref{doth}) can be integrated, from the very beginning, obtaining
\begin{equation} \label{hofa}
 h(a) = h(a_*) - 3 \int_{a_*}^a a'{}^2\pi(a') da'\,,
\end{equation}
where $a_*=a(T_*)$ and following the considerations
given in Sec.~\ref{Section:BackCosm}, $a_*$ must be the value of the scale factor at a very
early time. Using now Eq.~(\ref{rhoofTandn}), we get
\begin{equation} \label{rhoPh}
 \rho(a) = - P_0 - \frac{3}{a^3}  \int_{a_*}^a a'{}^2\pi(a') da' +
\frac{h(a_*)}{a^3}.
\end{equation}
The energy change, due to dissipative  bulk viscosity, gets the
attractive form $\Delta U =  -\pi(V) \Delta V$.
Although Eq.~(\ref{hofa}) is an exact solution for the case $\pi=\pi(a)$,
the general case obeys the same relation once
we found the solution $\pi(a) = \pi(a,\dot{a}(a))$. Given the freedom
on $\zeta$ and $\tau_0$
as functions of $\rho$ in Eqs. (\ref{BVET}), (\ref{ISTTE}) or
(\ref{ISTTET}), there are uncountable different forms that the
bulk viscosity can take. This enables to follow a phenomenological approach by employing
a particular dependence $\pi(a)$, which thereafter could be naively compared
with observations.

As a first example of this approach, let us consider the case in
which $\pi$ is solved as a power-law function of the scale factor,
\begin{equation}
 \pi = \pi_0 a^{\epsilon},
\end{equation}
where $\epsilon$ and $\pi_0<0$ are real constants.
The contribution to the energy density $\rho_{diss}$, due
to dissipative quantities, becomes then
\begin{equation}
 \rho_{diss} = \frac{3|\pi_0|}{\epsilon + 3}a^{\epsilon},  \qquad
(\epsilon \neq -3)
\end{equation}
where we consider $a-a_* \simeq a$.

As a result, in the case $\epsilon>0$, we find that $\rho_{diss}$ leads to a phantom
behavior. Thus, we may consider a time $t_i$ for which we neglect all contributions to
the energy
density except the dissipative quantities. It is therefore straightforward to show that the
scale factor will blow up to infinity in a finite time given by:
\begin{equation}
 t_{Big Rip} = t_i + \frac{\epsilon + 3}{8\pi G |\pi_0| \epsilon
a_i^{\epsilon/2}}\,,
\end{equation}
where $a_i \equiv a(t_i)$, corresponding to a {\it Big Rip} \cite{Caldwell:2003vq} (see also \cite{Astashenok:2012iy,Pavon:2012pt}).

\begin{figure}[ht]
	\begin{center}
	\includegraphics[width=3 in]{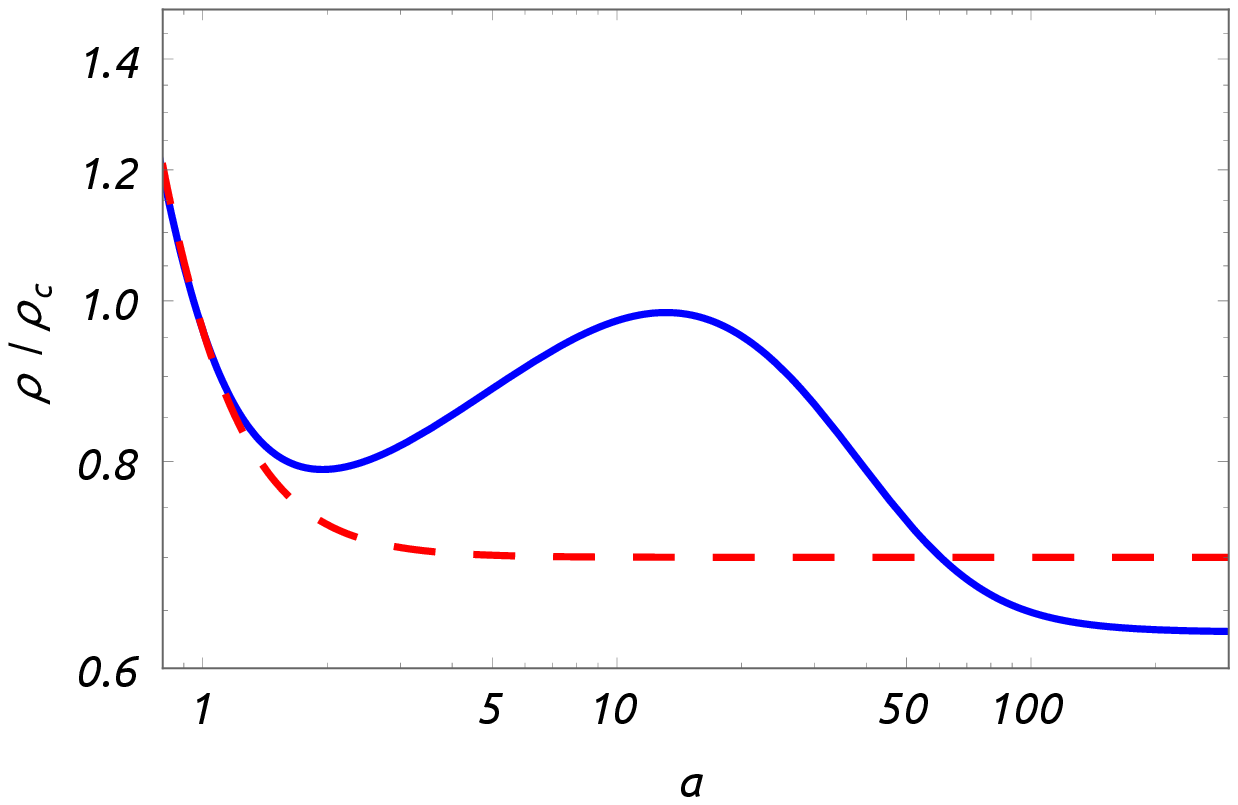}
	\includegraphics[width=3 in]{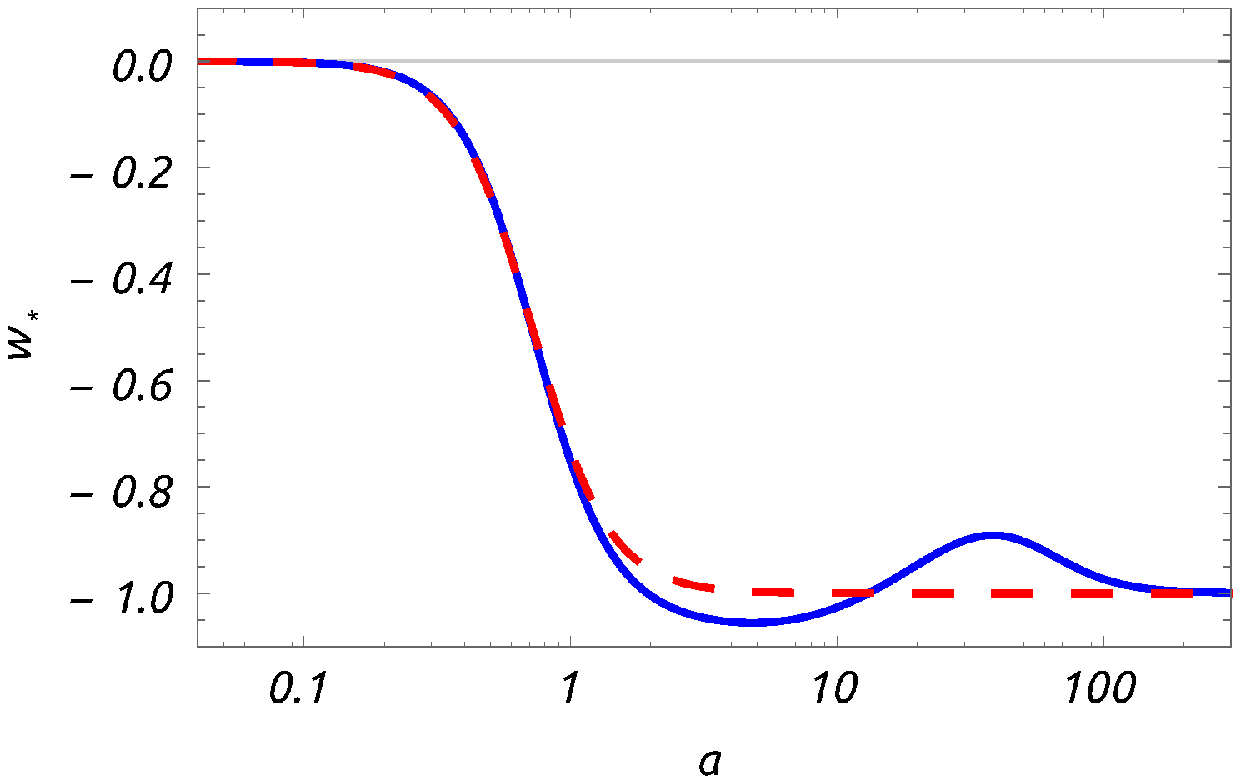}
	\caption{Evolution of the little rip model expressed by Eq.~(\ref{LRV}) (solid curves) and of the
	dark fluid (dashed curves). Left panel: Energy density in units of the critical energy
	density nowadays as a function of the scale factor. Right panel: Effective EoS parameter as a function of the scale factor.
	Here $a_P =10$ and $|\pi_0|=0.1 \rho_c$.
	\label{fig:LittleRip}}
	\end{center}
\end{figure}

A second interesting example shows up a Little Rip behavior
\cite{Frampton:2011sp,Brevik:2011mm,Frampton:2011rh,Astashenok:2012tv}. In our formalism this can be achieved, for example, by considering the function
\begin{equation} \label{LRV}
 \pi(a) = - |\pi_0| a e^{-a/a_P}\,,
\end{equation}
where $a_P$ is a given scale. Using Eqs.~(\ref{rhoPh}) and (\ref{LRV}), or by directly integrating Eq.~(\ref{ContEqwV}), we obtain
the energy density
\begin{equation}
 \rho(a) = -P_0 + \frac{h(a_*)}{a^3} + \rho_{diss}\,,
\end{equation}
where
\begin{equation} \label{LRExrhodiss}
 \rho_{diss} = \frac{18|\pi_0|a_P^4}{a^3} - 3|\pi_0|a_p e^{-a/a_P}
 \Big[1+ 3 \frac{a_P}{a} +6\left(\frac{a_P}{a}\right)^2 +6\left(\frac{a_P}{a}\right)^3\Big]\,,
\end{equation}
is the term induced by the dissipative viscosity of the dark fluid, in which again we have considered $a-a_* \simeq a$.
Further, we define the effective EoS parameter $w_*$ as
\begin{equation}
 w_* = \frac{-P_0 + \pi(a)}{\rho}\,.
\end{equation}
In Fig.~\ref{fig:LittleRip} we show the evolution of the energy density of the dissipative dark fluid and its EoS parameter. We note
that although $w_*<-1$ in a given interval of the scale factor, it tends towards $-1$ as the scale factor grows, leading to a Little Rip solution.
In this same interval we note that the energy density is a growing function of the scale factor. As a reference, we also show the energy density
of the dark fluid without dissipatives and its EoS parameter. Both models are adjusted so that their present abundances are $\Omega_d = 0.96$.

\subsection{The arbitrary decomposition and phantom behavior}

In Sec. I, we emphasized that the phantom behavior occurs
when decomposing the dark sector into two species. In this section we show explicitly how this
may occur in our formalism.

We decompose the dark fluid including dissipative terms in dark energy and dark matter contributions, as $\rho = \rho_{de} +
\rho_{dm}$, with
\begin{equation}
\rho_{dm} = \frac{h(a_*)}{a^{3}}\,,
\end{equation}
\begin{equation}
 \rho_{de} = -P_0 + \frac{3}{a^{3}}\int^{a}_{a_*} a'{}^{2} \pi(a')da'\,.
\end{equation}
Rephrasing it differently, we are interpreting the consequences of the out-of-equilibrium processes as a
dark energy effect. The condition for having a phantom dark energy is $\rho_{de} + P_0 + \pi < 0$, or alternatively
\begin{equation}
 a^{3}\pi(a) - 3 \int^{a}_{a_*} a'{}^{2} \pi(a')da' < 0\,,
\end{equation}
while the condition for the total dark fluid to be non-phantom is $\rho_{dm} + \rho_{de} + P_0 + \pi > 0$, or
\begin{equation}
 h(a_*) + a^{3} \pi(a) - 3 \int^{a}_{a_*} a'{}^{2} \pi(a')da' \geq 0\,,
\end{equation}
and leads also to a non-phantom dark energy. We can join those two conditions into a single one, having:
\begin{equation} \label{condForDFnp}
 0 <  3 \int^{a}_{a_*} a'{}^{2} \pi(a')da' - a^{3}\pi(a) \leq h(a_*)\,,
\end{equation}
which is equivalent to $0<-(\rho_{de} + P_0 + \pi) < \rho_{dm}$.

Thus, if dark fluid's dissipative effects accomplish the above inequality, we easily get that the decomposition 
into dark matter and dark energy provides a phantom dark energy contribution. This violates the dominant energy condition, 
whereas the dark fluid, as a whole, behaves non-phantom. As a simple example, we consider a constant $\pi_0$, the inequality (\ref{condForDFnp}) holds
if $0< -a^3_* \pi_0 < h(a_*)$. This particular example also shows that we may obtain the dark energy contribution from a pure
dark matter fluid, i.e. $P_0=0$, with dissipative processes.

\begin{figure}[ht]
	\begin{center}
	\includegraphics[width=3 in]{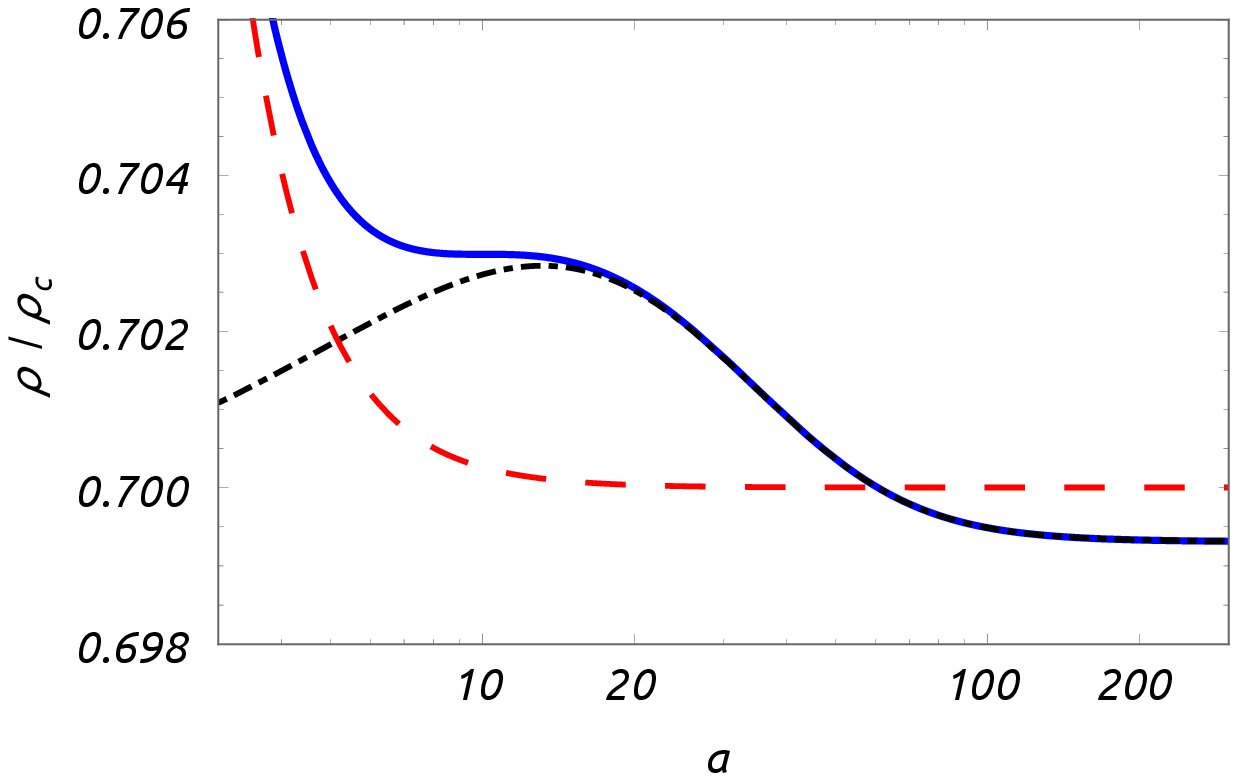}
	\includegraphics[width=3 in]{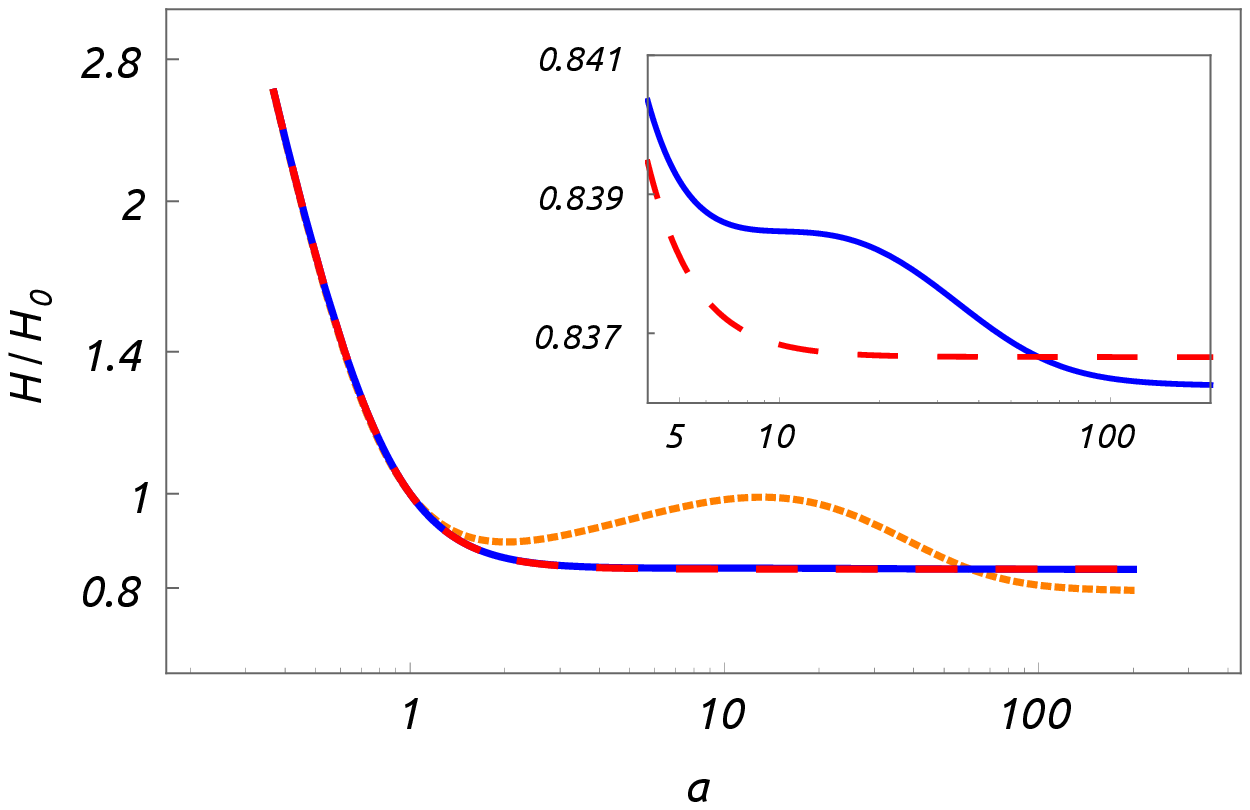}
	\caption{(Color online) Evolution for the little rip model expressed by Eq.~(\ref{LRV}). Left panel: Evolution of the energy density, 
	the dot-dashed curve shows
	the $\rho_{de}$ contribution and the solid curve (blue) the total dark fluid energy density. We note that, while the dark energy
	contribution is phantom in some region, the total dark fluid is not. Here $a_P =10$ and $|\pi_0|=0.001 \rho_c$. The long-dashed curve
	(red) shows the evolution of the dark fluid energy density without dissipatives. Right panel: Evolution of the Hubble factor. The dotted (orange)
	curve is for the model with $|\pi_0|=0.1$, corresponding to the case of Fig.~\ref{fig:LittleRip}, the other two curves as in the left panel.
	The inset shows a zoom to the region for which the differences between the cases of dark fluid with
	and without dissipatives are evident. We notice that for a
	viscosity strenght  $|\pi_0|=0.1 \rho_c$ there is a region for which $dH/da >0$, while for viscosity strenght $|\pi_0|=0.001 \rho_c$ there is not.
	\label{fig:LittleRip2}}
	\end{center}
\end{figure}

\noindent A more interesting example arises from the bulk viscosity defined in Eq.~(\ref{LRV}). We can decompose the whole dark
fluid as $\rho_d = \rho_{de} + \rho_{dm}$, with $\rho_{de} = -P_0 + \rho_{diss}$ ($\rho_{diss}$ has been given
by Eq.~(\ref{LRExrhodiss})) and $\rho_{dm} = h(a_*)/a^3$. For small values of the bulk viscosity strength $|\pi_0|$,
the total energy density $\rho_{d}$ does not become a growing function of the scale factor (contrary to the situation
in Fig.~\ref{fig:LittleRip}). In the left panel of Fig.~\ref{fig:LittleRip2}, we show such a behavior,
choosing $a_P = 10$ and $|\pi_0|=0.001 \rho_c$. The solid curve corresponds to the dark fluid energy density.
We soon note that $d\rho_d / da \leq 0$ in its whole domain, while
the arbitrary dark energy component, plotted with a dot-dashed curve, shows up a phantom behavior for $a \lesssim a_P$. It is worthly to note
that Eq.~(\ref{condForDFnp}) is accomplished in the interval where $d \rho_{de}/da \geq 0$. In the right panel of Fig.~\ref{fig:LittleRip2}
we show the evolution of the scale factor, the solid (blue) line corresponds to the this case, and we note that $H(a)$ is not growing, while
the case $|\pi_0|=0.1 \rho_c$, depicted with a dotted (orange) line is.

\subsection{Dissipative processes in cosmological perturbations} \label{Sect:CosmPert}

In this subsection, we give a brief description of the behavior of perturbations in the presence of dissipative processes. To this end, we write the metric of the perturbed universe for scalar perturbations in the Newtonian-conformal gauge as
\begin{equation}
 ds^2 = a^2(\tau)\big[ -(1+2 \Psi) d\tau^2 + (1- 2 \Phi) \delta_{ij} dx^i dx^j \,\big]. \label{pertmetric}
\end{equation}
The potentials $\Psi$ and $\Phi$ are the usual metric scalar perturbations, which are related to the Poisson equation and to the geodesic equations
respectively. We introduced the conformal time $\tau$ which, in the absence of perturbations, it is related
to the cosmic time $t$ by $d t = a(\tau) d\tau$. Using these coordinates the four-velocity becomes
\begin{equation}
 u^{\mu} = \frac{1}{a}\big( 1 - \Psi, v^i \big),
\end{equation}
where $v^i$ is the peculiar velocity, {\it i.e.}, the average velocity the particles have with respect to the Hubble flow.

In the coordinates described by Eq. (\ref{pertmetric}), the stress-energy tensor components become
\begin{eqnarray}
 T^{0}{}_{0} &=& -\bar{\rho} (1+\delta)\,, \\
 T^{0}{}_{i} &=& (1+w^*) \bar{\rho} v_i  + \frac{1}{a} q_i\,, \\
 T^{i}{}_{j} &=& (\bar{P} + \delta P + \pi + \delta \pi) \delta^{i}_{j} + \pi^{i}_{\,\,j}\,.
\end{eqnarray}
Note that $\bar{\rho}$, $\bar{P}$ and $\pi$ are background quantities which only depend on the time coordinate $\tau$, whereas
$\rho = \bar{\rho} + \delta \rho = \bar{\rho} (1+\delta)$, $P = \bar{P} + \delta P$ are respectively the total energy density and isotropic pressure of the fluid. The fluid expansion scalar becomes
\begin{equation}
 \Theta = \frac{1}{a} \Big( 3 \mathcal{H}(1-\Psi) +  \partial_i v^i - 3\dot{\Phi} \Big),
\end{equation}
and the shear viscosity is
\begin{equation}
 \sigma_{i j} =\frac{1}{2} a (\partial_i v_j + \partial_j v_i) - \frac{1}{3} a \partial_k v^k \delta_{ij}.
\end{equation}
We note that the vorticity does not appear in the linear perturbed equations, because it is multiplied
in Eq.~(\ref{EntProd:HydrodEulerEq}) by the heat transfer, which is also a perturbed quantity.
It is convenient to define $v_q^i$ through the components of the heat transfer $q^\mu$ as
\begin{equation}
 q^\mu = \frac{(1+w^*) \bar{\rho} }{a} (0, v^i_q).
\end{equation}
With this definition the indices of $v_{q}^i$ (as well as for $v^i$) are raised and lowered through the euclidian metric $\delta_{ij}$.
After some manipulations of Eqs.~(\ref{EntProd:ContEq}) and (\ref{EntProd:HydrodEulerEq}), we rearrange the continuity equation
\begin{equation} \label{CosmP:ContEq}
 \partial_\tau \delta + (1+w^*)\left(\theta^* - 3 \partial_\tau \Phi \right) +
 3 \mathcal{H}\left(\frac{\delta P^*}{\delta \rho} - w^* \right) \delta = 0,
\end{equation}
and the Euler equation
\begin{equation}\label{CosmP:EulerEq}
 \partial_\tau \theta^* + \mathcal{H} (1 - 3 w^*)\theta^* + \frac{\partial_\tau w^*}{1+w^*} \theta^*
 - k^2 \frac{ \delta P^*/\delta \rho}{1+w^*} \delta - k^2 \Psi  + k^2 \pi_s  = 0\,,
\end{equation}
which have been conventionally reported in the Fourier space. We also defined
\begin{equation}
 \theta^* \equiv \theta + \theta_q\,,
\end{equation}
where as usual $\theta \equiv \partial_i v^i$, and analogously: $\theta_q \equiv \partial_i v^i_q$.
We also get the shear viscosity scalar
\begin{equation}
 \pi_s \equiv -\frac{k_ik^j \pi^{i}_{\,j}}{ k^2 (1+w^*)\bar{\rho}} = \frac{4 \eta \theta}{3 a (1+w^*)\bar{\rho}}.
\end{equation}
(Note that contrary to $k^i$, $\pi_{ij}$ are components of a tensor.) As usual, we chose the shear viscosity to satisfy
$\pi_{\mu\nu} = - 2 \eta \sigma_{\mu\nu}$, ensuring that the entropy flux grows as it can be seen in
Eq.~(\ref{EntProd:EntropyFlux}). We notice that apart from the scalar shear viscosity in Eq.~(\ref{CosmP:EulerEq}), the continuity and Euler equations are
the same of a perfect fluid with pressure $P_*=P_0 + \pi$. We further require the use of Einstein's equations, which give
\begin{equation} \label{CosmP:PoissonEq}
 k^2 \Phi = -4 \pi G a^2  \rho \Delta^{*},  \qquad \text{where} \qquad \Delta^{*} \equiv \delta + 3 \mathcal{H} (1+w^{*}) \frac{\theta^{*}}{k^2},
\end{equation}
and
\begin{equation} \label{CosmP:ShearEq}
 k^2( \Phi - \Psi) = 16 \pi G a^2 \, \eta \frac{\theta}{a},
\end{equation}
where we perform a sum also over other non-dark fluid contributions to the right hand side of the above equations. Immediately, a relevant consequence arises: if shear viscosity is here neglected, this set of equations does not allow one to distinguish heat transfer from  dark fluid's velocity divergence. Rephrasing it differently, one gets that only bulk viscosity can be constrained by linear perturbation theory. However, this opens the possibility to use observations of the visible matter velocities, at non-linear orders, in order to isolate $\theta$ from $\theta_*$, gaining valuable information on dark fluid's heat transfer. This treatment is beyond the aims of this paper, but will be object of future investigations. In the following, we therefore neglect shear viscosity and solve the perturbed system in the presence of bulk viscosity. For completeness, we consider as specific model the one given by Eq.~(\ref{LRV}) with $a_P = 10$ and $|\pi_0| = 0.001 \rho_c$. Given these values, the dark fluid is non-phantom and $1+w_* >0$ at all stages of its evolution. In so doing, the Euler equation, as reported in Eq.~(\ref{CosmP:EulerEq}), is also well defined. To obtain the perturbation of the viscosity $\pi(a)$, we make use of the substitution $a \rightarrow a(1+\Psi)$, having
\begin{displaymath}
 \delta \pi = \frac{a}{a_P}|\pi_0| (a-a_P) e^{-a/a_P} \Psi.
\end{displaymath}

We may describe the evolutions of Eqs.~(\ref{CosmP:ContEq}), (\ref{CosmP:EulerEq}), (\ref{CosmP:PoissonEq}), and the corresponding continuity
and Euler equations for baryons. To do so, we notice that the differences with the $\Lambda$CDM model arise only at late times. 
At that epoch, we neglect the contribution of radiation and we start the evolution at redshifts circumscribed after the recombination.

\begin{figure}[ht]
	\begin{center}
	\includegraphics[width=3 in]{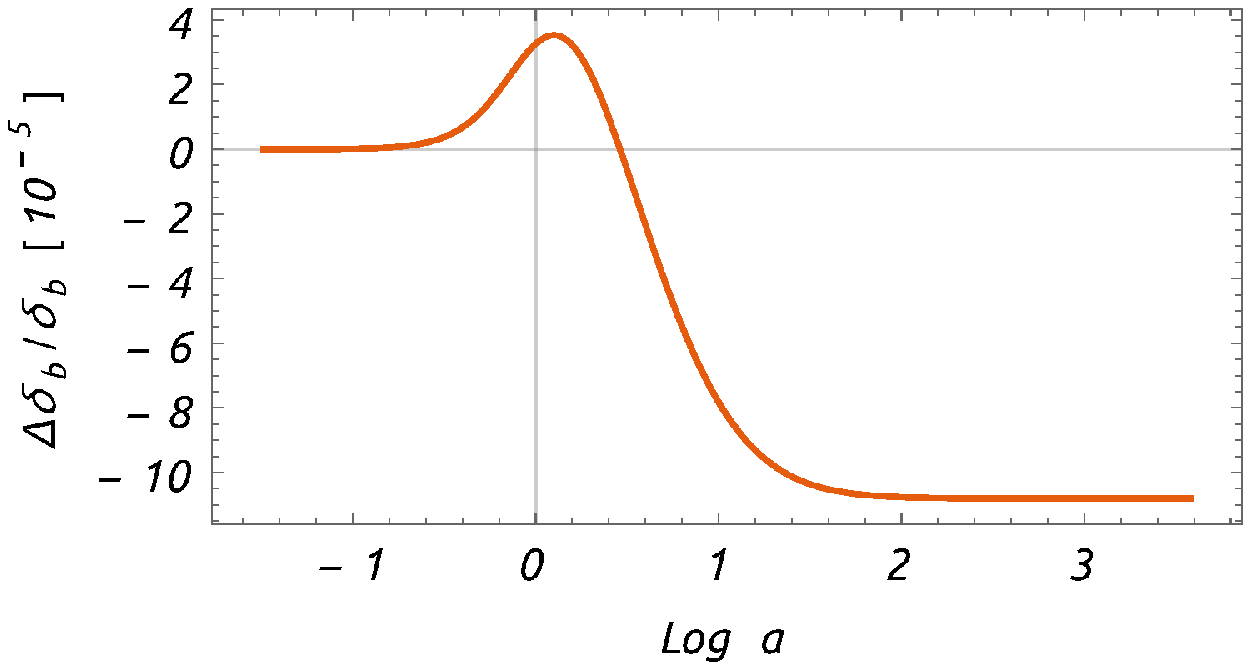}
	\includegraphics[width=3 in]{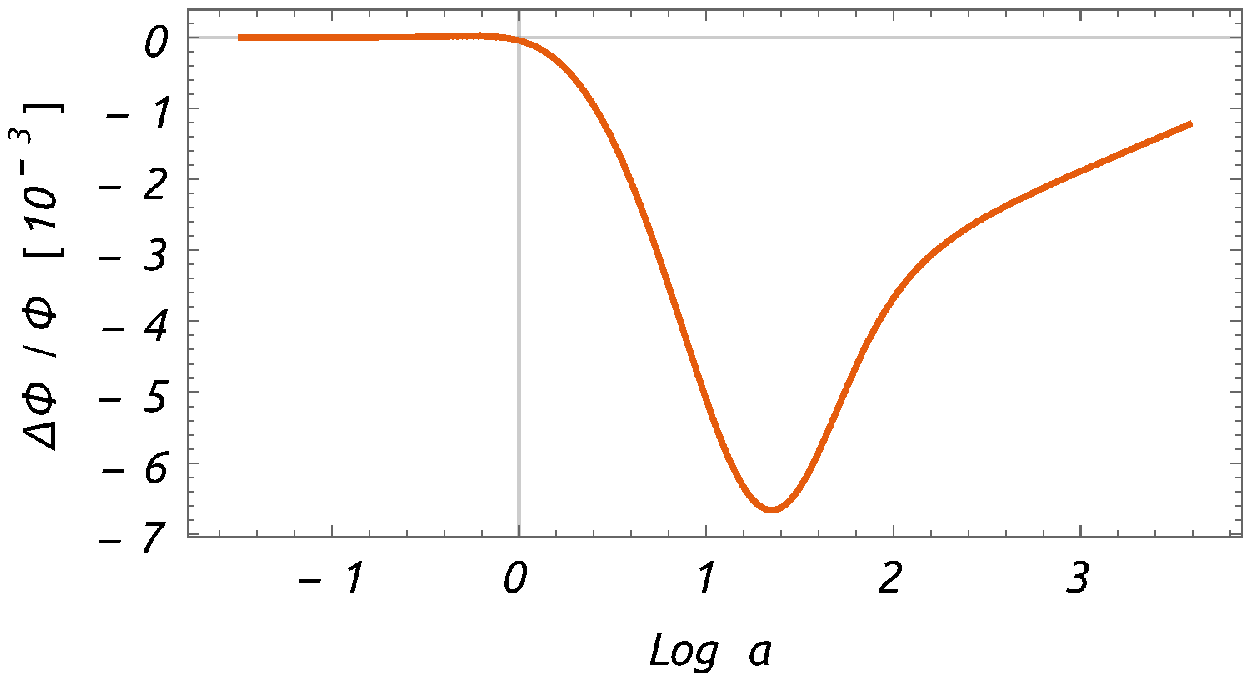}
	\caption{Perturbation evolutions of the little rip model, defined in Eq.~(\ref{LRV}), evolving as a function of the scale factor 
	logarithm. In particular, in the left panel we plot the differences in the evolution of the energy density contrast of baryonic matter, 
	while in the right panel we plot the difference in the evolution of the Newtonian potential $\Phi$.	\label{pert}}
	\end{center}
\end{figure}

In Figs.~\ref{pert} we show the differences between the model with viscosity and the pure dark fluid.
We plot the differences
\begin{displaymath}
\frac{ \Delta \delta_b}{\delta_b} = \frac{\delta_b-\delta_b^{(fid)}}{\delta_b}, \qquad
\frac{ \Delta \phi_b}{\phi_b} = \frac{\phi_b-\phi_b^{(fid)}}{\phi_b},
\end{displaymath}
where the label $fid$ refers to as the fiducial dark fluid without dissipative terms (degenerated with the
$\Lambda$CDM model).

\section{Final outlooks and perspectives} \label{Sect:Conclusions}

In this work, we investigated the thermodynamical properties of a class of fluids whose corresponding pressure is a non-zero constant. In the context
of cosmology this restriction gives rise to a unified dark energy fluid, for which both dark energy and dark matter emerge as a single effect. 
This \emph{dark fluid hypothesis} lies on the assumption that the involved fluid is barotropic and provides a vanishing speed of sound. 
The latter property allows perturbations to grow at all scales and at the same time provides a constant negative pressure that accelerates 
the universe. Previous investigations on the dark fluid were mainly based on assuming the validity of the continuity equation, valid for 
pure adiabatic processes. Consequently, the characteristic scale $\mathcal{K}$ would represent an integration constant, as here showed, 
and cannot be physically well-supported in order to describe the difference between the dark fluid and the standard $\Lambda$CDM model. 
In this paper, we studied the dark fluid properties in a more general thermodynamical context. We therefore showed that its evolution 
generally differs from the one associated to the standard $\Lambda$CDM model. In particular, we found that $\mathcal{K}$ is no longer 
an integration constant, becoming instead a function of the temperature, described by the ratio of  fluid's enthalpy and pressure 
contained in a unitary co-moving volume. The consequences in cosmology are relevant. First, we demonstrated that, for constant 
pressure systems, adiabatic and isothermal processes should coincide, and therefore the dark fluid and the $\Lambda$CDM models 
degenerated in the case of adiabatic expansions only. Given that fluids are generally non-perfect, we expected that dissipative 
processes arose, leading to an important breaking of the above mentioned degeneracy between the standard model and the dark fluid itself. 
Thus, we treated the case of bulk viscosity in detail at the background and perturbed cosmological levels and we also considered to 
introduce the shear viscosity and heat transfer. We followed a phenomenological approach and from the beginning we gave a dependence on 
the scale factor to the bulk viscosity. Our treatment deeply departures from either Eckart's or MIS theories. Afterwards, we explicitly 
showed that our working scheme is somehow equivalent to Eckart's theory, providing similar outcomes. However, since Eckart's and MIS 
theories have the disadvantage not to fix their free functions, we concluded that our approach better motivated the physics behind the 
phenomenological description of the dark fluid in constant pressure systems. In fact, one important aspect we addressed in our work was 
to describe a phantom dark energy behavior. It is remarkable to notice, in fact, that cosmological observations, albeit consistent with 
a non-evolving dark energy component, do not exclude \emph{a priori} an EoS parameter lesser than 1 and therefore we even took into 
account this possibility. In any cases, theoretical problems jeopardize the existence of a phantom dark energy term. Thus, we argued that, 
if one assumes phantom dark energy, it would represent some sort of signature for a unified fluid, instead of a dominant energy condition 
violating component. In other words, we demonstrated that it is possible to simply infer a phantom behavior as a natural consequence of 
the dark fluid existence. To this end, we showed by means of a ``little rip'' toy model how this behavior can be easily recovered. 
Since our analysis is almost independent from  the dissipative processes involved into the treatment, it is straightforward to imagine 
a possible generalization to more complex situations. For example, one may suppose the case of the complex scalar field from which some 
sort of generalized Chaplygin gas emerges. We concluded our work with explicitly writing the dark fluid perturbation equations in the 
presence of dissipative processes. Further, we showed that in the absence of stress viscosity, it is not possible to distinguish between 
the particle velocities and heat flows with only keeping in mind those equations. Finally, we showed the dynamics of a bulk viscosity term, 
leading to a little rip evolution. Future 
efforts will be devoted to test our model with present time data, in order to distinguish our model from the standard $\Lambda$CDM 
paradigm. In addition, we will extend our formalism by means of higher order perturbation theory and we will try to better alleviate 
the degeneracy between the $\Lambda$CDM model and our picture.

\section{Aknowledgements}
A.A. and O.L. want to thank prof. S.~Capozziello and G.~Carmona for useful discussions. A.A.
acknowledges the hospitality of the Departamento de F\'isica, Universidad de Santiago de Chile, where part of this work was done.
A.A. and J.K. are financially supported by the project CONACyT-EDOMEX-2011-C01-165873 (ABACUS-CINVESTAV).
N.C. acknowledges the support to this research by CONICYT through grants Nos. 1140238.
O.L. is financially supported by the European PONa3 00038F1 KM3NeT (INFN) Project.

\appendix*

\section{Exact solutions in Eckart's theory} \label{appendixA}

Exact cosmological soulutions in Eckart's theory can be obtained for the dark fluid when the viscosity
has a power-law dependence upon the energy density of this fluid
\begin{eqnarray}\label{xilaw}
\xi=\xi_{0} \rho^{m},
\end{eqnarray}
where $\xi_{0}>0$ and $m$ are constant parameters. This type of behavior for the viscosity
has been widely investigated in the literature, albeit there is no fundamental complete 
approaches for choosing it, see for example Ref.~\cite{Maartens:1995wt}. We will assume this type of behavior which allows us
to obtain suitable cosmological solutions and compare with other results present in the literature.
Neglecting all contributions to the total energy momentum tensor, except the dark fluid, we can write down
\begin{equation}
 \pi(\rho) = -3 H \xi_{0} \rho^{m} = - \sqrt{3} \sqrt{8\pi G} \xi_{0} \rho^{m + 1/2}\,.
\end{equation}
Only for reasons of mathematical simplicity the case $m=1/2$ is mostly considered.  As a first
glance to the study of the behavior of this fluid when dissipation is taken into account, it is reasonable to explore
this simple case. Thus, in the following
\begin{equation} \label{A:piofrho}
 \pi(\rho) = -\sqrt{3} \tilde{\xi}_0 \rho,
\end{equation}
where we define $\tilde{\xi}_0 \equiv \sqrt{8\pi G} \xi_{0}$.
The continuity equation becomes
\begin{equation} \label{A:Conteq}
 \dot{\rho} + 3 H (\alpha \rho + P_0) = 0,
\end{equation}
where $\alpha \equiv 1-\sqrt{3}\tilde{\xi}_{0}$.
This can be integrated to get the energy density as a function of the scale factor:
\begin{equation} \label{A:rhoofa}
 \rho = - \frac{P_0}{\alpha} + \left(\rho_0 + \frac{P_0}{\alpha} \right) a^{-3\alpha}.
\end{equation}
Using Eq.~(\ref{A:piofrho}) we obtain $\pi$ as a function of the scale factor
\begin{equation} \label{A:piofa}
 \pi(a) = -\sqrt{3} \tilde{\xi}_0 \left[ - \frac{P_0}{\alpha} + \left(\rho_0 + \frac{P_0}{\alpha} \right) a^{-3\alpha} \right]
\end{equation}
We notice that Eq.~(\ref{rhoPh}), and hence Eq.~(\ref{BackCosm:energydensity}), is recovered by
\begin{displaymath}
 h(a_*) = \left(\rho_0 + \frac{P_0}{\alpha} \right) a_*^{3(1-\alpha)} - \frac{\sqrt{3} \tilde{\xi}_0 P_0}{1-\sqrt{3}\tilde{\xi}_0} a^3_*
\end{displaymath}

To study this solution in more detail, we can combine the Friedmann and continuity equations to obtain
\begin{equation}\label{Hpuntounmedio}
2\dot{H}+ 3 \alpha H^{2} + 8 \pi G P_{0} =0.
\end{equation}
The case $\alpha >0$ can be integrated to give
\begin{equation}\label{Halfamay0}
H(t)=\sqrt{-\frac{8 \pi G P_{0}}{3\alpha}}\left[\frac{e^{ \sqrt{- 24 \pi G P_{0}\alpha}\,t}  - A}{ e^{\sqrt{- 24 \pi G P_{0}\alpha}\,t}  + A }                        \right]  ,
\end{equation}
where $A$ is defined by
\begin{equation}\label{A}
A \equiv e^{ \sqrt{-24 \pi G P_{0}\alpha}\,t_0}\frac{\sqrt{\frac{- 8\pi G P_{0}}{3\alpha}}-H_{0}}{\sqrt{\frac{-8\pi G P_{0}}{3\alpha}}+H_{0}}.
\end{equation}
On the contrary, the case $\alpha < 0$ is obtained by analytic continuation of Eq.~(\ref{Halfamay0}), which turns out to be a real function of $t$.
We notice that the solution behaviors are strongly dependent upon the sign of the parameter $\alpha$.

\bibliographystyle{natbib}  
\bibliography{bibliography}  

\end{document}